\documentclass{elsart}
\usepackage{graphicx}
\usepackage{amssymb}
\usepackage{amsmath}

\begin{document}
\newcommand{\aff}[2]{Dipartimento di Fisica dell'Universit\`a 
  #1 e Sezione INFN, #2, Italy.}
\newcommand{\affuni}[2]{Dipartimento di Fisica dell'Universit\`a 
  #1, #2, Italy.}
\newcommand{\affinfn}[2]{INFN Sezione di #1, #2, Italy.}
\newcommand{\affinfnm}[2]{INFN Sezione di #2, #2, Italy.}
\newcommand{\affinfnn}[2]{INFN Sezione di #1, #2, Italy.}
\newcommand{\affd}[1]{Dipartimento di Fisica dell'Universit\`a 
  e Sezione INFN, #1, Italy.}

\newcommand{\dafne}	{DA$\Phi$NE }

\newcommand{\phirpi}	{\phi \rightarrow \rho \pi}
\newcommand{\phikk}	{\phi \rightarrow K \bar{K}}
\newcommand{\phikkrpi}	{\phi \rightarrow K \bar{K} / \rho \pi}
\newcommand{\phippp}	{\phi \rightarrow \pi^+ \pi^- \pi^0}
\newcommand{\phietag}	{\phi \rightarrow \eta \gamma}
\newcommand{\phietaee}	{\phi \rightarrow \eta \, e^+ e^- }

\newcommand{\etappeeg}	{\eta \rightarrow \pi^+ \pi^- e^+ e^- (\gamma)}
\newcommand{\etappee}	{\eta \rightarrow \pi^+ \pi^- e^+ e^- }
\newcommand{\etaeeeeg}	{\eta \rightarrow e^+ e^- e^+ e^- (\gamma)}
\newcommand{\etaeeee}	{\eta \rightarrow e^+ e^- e^+ e^- }
\newcommand{\pp}	{\pi^+ \pi^- }
\newcommand{\ee}	{e^+ e^- }
\newcommand{\ppee}	{\pi^+ \pi^- e^+ e^- }
\newcommand{\eeee}	{e^+ e^- e^+ e^- }

\newcommand{\etapp}	{\eta \rightarrow \pi^+ \pi^-}
\newcommand{\etappg}	{\eta \rightarrow \pi^+ \pi^- \gamma}
\newcommand{\etaeeg}	{\eta \rightarrow e^+ e^- \gamma}
\newcommand{\etagg}	{\eta \rightarrow \gamma \gamma}
\newcommand{\etappp}	{\eta \rightarrow \pi^+ \pi^- \pi^0}
\newcommand{\etapOpOpO}	{\eta \rightarrow \pi^0 \pi^0 \pi^0}

\newcommand{\bharad}	{e^+ e^- \to e^+ e^- (\gamma)}
\newcommand{\eephietag}	{e^+ e^- \to \phi \to \eta \gamma}

\newcommand{\aphi}	{\mathcal{A}_{\phi}}
\newcommand{\spcp}	{\sin \phi \cos \phi}
\newcommand{\spcpp}	{\sin \phi \cos \phi>0}
\newcommand{\spcpm}	{\sin \phi \cos \phi<0}
\newcommand{\ltb}       {\langle \cos \theta_b \rangle}
\newcommand{\ltf}       {\langle \cos \theta_f \rangle}
\newcommand{\sfp}       {\sum_1^4 |\vec{p}_i|}

\begin{frontmatter}

\title{Observation of the rare 
  \mathversion{bold}$\etaeeee$\mathversion{normal} decay with the KLOE
experiment}

\collab{The KLOE Collaboration}
\author[Na,infnNa]{F.~Ambrosino},
\author[Frascati]{A.~Antonelli},
\author[Frascati]{M.~Antonelli},
\author[Roma2,infnRoma2]{F.~Archilli},
\author[Cracow]{I.~Balwierz},
\author[Frascati]{G.~Bencivenni},
\author[Roma1,infnRoma1]{C.~Bini},
\author[Frascati]{C.~Bloise},
\author[Roma3,infnRoma3]{S.~Bocchetta},
\author[Frascati]{F.~Bossi},
\author[infnRoma3]{P.~Branchini},
\author[Frascati]{G.~Capon},
\author[Frascati]{T.~Capussela},
\author[Roma3,infnRoma3]{F.~Ceradini},
\author[Frascati]{P.~Ciambrone},
\author[Frascati]{E.~Czerwi\'nski},
\author[Frascati]{E.~De~Lucia},
\author[Roma1,infnRoma1]{A.~De~Santis},
\author[Frascati]{P.~De~Simone},
\author[Roma1,infnRoma1]{G.~De~Zorzi},
\author[Mainz]{A.~Denig},
\author[Roma1,infnRoma1]{A.~Di~Domenico},
\author[infnNa]{C.~Di~Donato},
\author[Roma3,infnRoma3]{B.~Di~Micco},
\author[Frascati]{M.~Dreucci},
\author[Frascati]{G.~Felici},
\author[Roma1,infnRoma1]{S.~Fiore},
\author[Roma1,infnRoma1]{P.~Franzini},
\author[Frascati]{C.~Gatti},
\author[Roma1,infnRoma1]{P.~Gauzzi},
\author[Frascati]{S.~Giovannella\corauthref{cor}},
\ead{simona.giovannella@lnf.infn.it}
\author[infnRoma3]{E.~Graziani},
\author[Frascati]{M.~Jacewicz},
\author[Frascati,StonyBrook]{J.~Lee-Franzini},
\author[Moscow]{M.~Martemianov},
\author[Frascati,Energ,Marconi]{M.~Martini},
\author[Na,infnNa]{P.~Massarotti},
\author[Na,infnNa]{S.~Meola},
\author[Frascati]{S.~Miscetti},
\author[Frascati]{G.~Morello},
\author[Frascati]{M.~Moulson},
\author[Mainz]{S.~M\"uller},
\author[Na,infnNa]{M.~Napolitano},
\author[Roma3,infnRoma3]{F.~Nguyen},
\author[Frascati]{M.~Palutan},
\author[infnRoma3]{A.~Passeri},
\author[Frascati,Energ]{V.~Patera},
\author[Roma3,infnRoma3]{I.~Prado~Longhi},
\author[Frascati]{P.~Santangelo},
\author[Frascati]{B.~Sciascia},
\author[Cracow]{M.~Silarski},
\author[Frascati]{T.~Spadaro},
\author[Roma3,infnRoma3]{C.~Taccini},
\author[infnRoma3]{L.~Tortora},
\author[Frascati]{G.~Venanzoni},
\author[Frascati,Energ,CERN]{R.~Versaci\corauthref{cor}},
\ead{roberto.versaci@lnf.infn.it}
\corauth[cor]{Corresponding author.}
\author[Frascati,Beijing]{G.~Xu},
\author[Cracow]{J.~Zdebik}

\collab{and, as members of the KLOE-2 collaboration:}
\author[Frascati]{D.~Babusci},
\author[Roma2,infnRoma2]{D.~Badoni},
\author[infnRoma1]{V.~Bocci},
\author[Roma3,infnRoma3]{A.~Budano},
\author[Moscow]{S.~A.~Bulychjev},
\author[Frascati]{P.~Campana},
\author[Frascati]{E.~Dan\'e},
\author[INFNBari]{G.~De~Robertis},
\author[Frascati]{D.~Domenici},
\author[Bari,INFNBari]{O.~Erriquez},
\author[Bari,INFNBari]{G.~Fanizzi},
\author[Roma2,infnRoma2]{F.~Gonnella},
\author[Frascati]{F.~Happacher},
\author[Uppsala]{B.~H\"oistad},
\author[Energ,Frascati]{E.~Iarocci},
\author[Uppsala]{T.~Johansson},
\author[Moscow]{V.~Kulikov},
\author[Uppsala]{A.~Kupsc},
\author[INFNBari]{F.~Loddo},
\author[Moscow]{M.~Matsyuk},
\author[Roma2,infnRoma2]{R.~Messi},
\author[infnRoma2]{D.~Moricciani},
\author[Cracow]{P.~Moskal},
\author[INFNBari]{A.~Ranieri},
\author[Frascati]{I.~Sarra},
\author[Calabria,INFNCalabria]{M.~Schioppa},
\author[Energ,Frascati]{A.~Sciubba},
\author[Warsaw]{W.~Wi\'slicki},
\author[Uppsala]{M.~Wolke}


\address[Frascati]{Laboratori Nazionali di Frascati dell'INFN, 
Frascati, Italy.}
\address[Cracow]{Institute of Physics, Jagiellonian University,
  Krakow, Poland.} 
\address[Mainz]{Institut f\"ur Kernphysik, Johannes Gutenberg -
  Universit\"at Mainz, Germany.} 
\address[Na]{Dipartimento di Scienze Fisiche dell'Universit\`a 
``Federico II'', Napoli, Italy}
\address[infnNa]{INFN Sezione di Napoli, Napoli, Italy}
\address[Energ]{Dipartimento di Scienze di Base ed Applicate per
  l'Ingegneria dell'Universit\`a ``Sapienza'', Roma, Italy.}
\address[Roma1]{\affuni{``Sapienza''}{Roma}}
\address[infnRoma1]{\affinfnm{``Sapienza''}{Roma}}
\address[Roma2]{\affuni{``Tor Vergata''}{Roma}}
\address[infnRoma2]{\affinfnn{Roma Tor Vergata}{Roma}}
\address[Roma3]{\affuni{``Roma Tre''}{Roma}}
\address[infnRoma3]{\affinfnn{Roma Tre}{Roma}}
\address[StonyBrook]{Physics Department, State University of New York
  at Stony Brook, USA.} 
\address[Beijing]{Institute of High Energy 
Physics of Academica Sinica,  Beijing, China.}
\address[Moscow]{Institute for Theoretical 
and Experimental Physics, Moscow, Russia.}
\address[Marconi]{Present Address: Dipartimento di Scienze e
  Tecnologie Applicate, 
Universit\`a Guglielmo Marconi, Roma, Italy.}
\address[CERN]{Present Address: CERN, CH-1211 Geneva 23, Switzerland.}
\begin{center}
and
\end{center}
\address[Bari]{\affuni{di Bari}{Bari}}
\address[INFNBari]{\affinfn{Bari}{Bari}}
\address[Calabria]{\affuni{della Calabria}{Cosenza}}
\address[INFNCalabria]{INFN Gruppo collegato di Cosenza, Cosenza, Italy.}
\address[Uppsala]{Department of Nuclear and Particle Physics, Uppsala
  Univeristy,Uppsala, Sweden.} 
\address[Warsaw]{A. Soltan Institute for Nuclear Studies, Warsaw, Poland.}

\begin{abstract}
We report the first observation of the rare $\etaeeee$ decay based on
1.7 fb$^{-1}$ collected by 
the KLOE experiment at the \dafne $\phi$-factory. 
The selection of the $\eeee$ final state is fully inclusive of radiation.
We have identified $362 \pm 29$ events resulting in a branching ratio of
$(2.4 \pm 0.2_{stat+bckg}\pm 0.1_{syst}) \times 10^{-5}$.
\end{abstract}

\end{frontmatter}

\section{Introduction}
\label{sec:introduction}
The $\etaeeee$ decay proceeds through two virtual photons 
intermediate state with internal photon conversion to $\ee$ pairs.
Conversion decays offer the possibility to precisely measure the
virtual photon 4-momentum via the invariant mass of the $\ee$ pair.
The lack of hadrons among the decay products makes the matrix element
directly sensitive to the $\eta$ meson transition form factor
\cite{Landsberg}.
The knowledge of the $\eta$ meson coupling to virtual photons
is important for the calculation of the anomalous magnetic moment
of the muon, being pseudoscalar exchange the major contribution to the 
hadronic light-by-light scattering.
\\
The first theoretical evaluation, 
$\Gamma(\etaeeee) / \Gamma(\etagg) = 6.6 \times 10^{-5}$,
dates from 1967 \cite{Jarlskog}.
The width ratio translates into a branching ratio (BR)
$BR(\etaeeee) = 2.59 \times 10^{-5}$ when the world average of the 
$BR(\etagg)$ measurements \cite{pdg2010} is taken as normalization
factor.
Other predictions exist in literature
\cite{Miyazaki:1974qi,Bijnens:1999jp,Lih:2009np,Petri:2010ea}, with
differences at the level of 10\%.
\\
Double lepton-antilepton $\eta$ decays have been searched by
the CMD-2 and the WASA experiments, obtaining the upper limits at 90\%
C.L., $BR(\etaeeee) < 6.9 \times 10^{-5}$ \cite{Akhmetshin:2000bw} and  
$BR(\etaeeee) < 9.7 \times 10^{-5}$ \cite{Berlowski:2008zz},
respectively.

\section{The KLOE detector}
\label{sec:detector}
The KLOE experiment operates at \dafne, the Frascati $\phi$-factory. 
DA$\Phi$NE is an $e^+e^-$ collider running at a center of mass energy 
of $\sim 1020$~MeV, the mass of the $\phi$ meson. 
Equal energy positron and electron beams collide at an angle 
of $\pi$-25 mrad, producing nearly at rest $\phi$ mesons.
\\
The detector consists of a large cylindrical Drift Chamber (DC),
surrounded by a lead-scintillating fiber electromagnetic calorimeter.
A superconducting coil around the EMC provides a 0.52~T field.
The beam pipe at the interaction region is spherical in shape  with 10
cm radius, it is made of a Beryllium-Aluminum alloy of 0.5 mm
thickness.
Low beta quadrupoles are located at about $\pm$50 cm distance from the
interaction region.
The drift chamber~\cite{DCH}, 4~m in diameter and 3.3~m long, has 12,582
all-stereo tungsten sense wires and 37,746 aluminum field wires. 
The chamber shell is made of carbon fiber-epoxy composite with an
internal wall of 1.1 mm thickness, 
the gas used is a 90\% helium, 10\% isobutane mixture. 
The spatial resolutions are $\sigma_{xy} \sim 150\ \mu$m and 
$\sigma_z \sim$~2 mm.
The momentum resolution is $\sigma(p_{\perp})/p_{\perp}\approx 0.4\%$.
Vertices are reconstructed with a spatial resolution of $\sim$ 3~mm.
The calorimeter~\cite{EMC} is divided into a barrel and two endcaps, for a
total of 88 modules, and covers 98\% of the solid angle. 
The modules are read out at both ends by photomultipliers, both in
amplitude and time. 
The readout granularity is $\sim$\,(4.4 $\times$ 4.4)~cm$^2$, for a total
of 2440 cells arranged in five layers. 
The energy deposits are obtained from the signal amplitude while the
arrival times and the particles positions are obtained from the time
differences. 
Cells close in time and space are grouped into calorimeter clusters. 
The cluster energy $E$ is the sum of the cell energies.
The cluster time $T$ and position $\vec{R}$ are energy-weighted averages. 
Energy and time resolutions are $\sigma_E/E = 5.7\%/\sqrt{E\ {\rm(GeV)}}$ 
and  $\sigma_t = 57\ {\rm ps}/\sqrt{E\ {\rm(GeV)}} \oplus100\ {\rm ps}$, 
respectively.
The trigger \cite{TRG} uses both calorimeter and chamber information.
In this analysis the events are selected by the calorimeter trigger,
requiring two energy deposits with $E>50$ MeV for the barrel and $E>150$
MeV for the endcaps. 
A cosmic veto rejects events with at least two energy deposits above 30 MeV
in the outermost calorimeter layer. 
Data are then analyzed by an event classification filter \cite{NIMOffline},
which selects and streams various categories of events in different
output files.

\section{Event selection}
\label{sec:eventselection}
The analysis has been performed using 1,733 pb$^{-1}$ from the
2004-2005 data set at $\sqrt{s} \simeq 1.02$ GeV.
242 pb$^{-1}$ of data taken off-peak at $\sqrt{s}=1.0$ GeV were
used to study the $\ee$ continuum.
Monte Carlo (MC) events are used to simulate the signal and the
background. 
The signal is generated according to the matrix element in
\cite{Bijnens:1999jp}, assuming $BR = 2.7 \times 10^{-5}$, in a sample 
of 167,531 pb$^{-1}$. 
Other MC samples are: 
  3,447 pb$^{-1}$ simulating the main $\phi$ decays
                  ($\phikk$ and $\phirpi$) and 
 17,517 pb$^{-1}$ simulating others more rare $\phi$ decays.
All MC productions account for run by run variations of the main
data-taking parameters such as background conditions, detector
response and beam configuration.
Data-MC corrections for calorimeter cluster energies and tracking efficiency,
evaluated with radiative Bhabha events and $\phirpi$ samples respectively,
have been applied.
Effects of Final State Radiation (FSR) have been taken into account using
the PHOTOS MC package \cite{Barberio:1993qi,Golonka:2005pn}.
This package simulates the emission of FSR photons by any of the decay
products taking also into account the interference between different
diagrams.
PHOTOS is used in the Monte Carlo at the event generation
level, so that our simulation fully accounts for radiative effects.
\\
At KLOE, $\eta$ mesons are produced together with a monochromatic recoil
photon ($E_{\gamma} = 363$ MeV) through the radiative decay $\phietag$.
In the considered data sample about $72 \times 10^{6}\ \eta$'s are
produced.
As first step of the analysis, a preselection is performed
requiring at least four (two positive and two negative) tracks
extrapolated inside a fiducial volume defined by a cylinder centered
in the interaction point and having radius $R = 4$ cm and length
$\Delta z = 20$ cm. 
For each charge, the two tracks with the highest momenta are selected.
One and only one neutral cluster, having energy
$E_{cl} \ge 250$ MeV and polar angle in the range $(23^\circ -157^\circ)$, 
is required. 
A cluster is defined neutral if it does not have any associated track
and has a time of flight compatible with the photon hypothesis.
To improve the energy and momentum resolution, a kinematic fit is performed
imposing the four-momentum conservation and the photon time of flight.
A very loose cut on the $\chi^2$ of the kinematic fit ($\chi^2<4000$)
is applied in order to discard poorly reconstructed events.

\section{Background rejection}
\label{sec:backgroundrejection}
\begin{figure}
  \begin{center}
    \includegraphics*[width=0.49\textwidth]{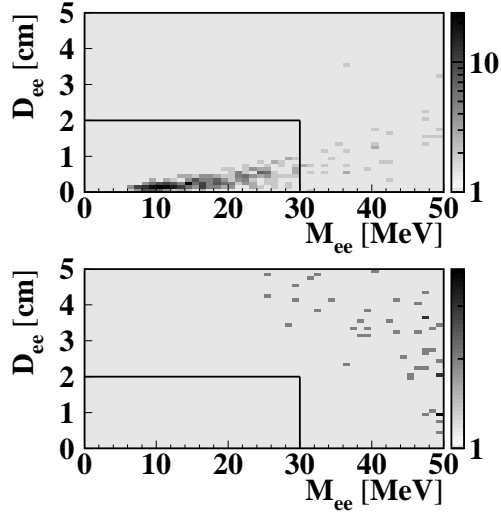}
  \end{center}
  \caption{$D_{ee}$ vs $M_{ee}$ evaluated at the drift chamber wall
           for MC $\phietag$ background (top panel) and MC signal
           (bottom panel).
           Events in the box 
           $M_{ee}(DCW) < 30\ {\rm MeV}\ \cap\ D_{ee}(DCW) < 2$ cm are
           rejected.}
  \label{fig:gee}
\end{figure}

\begin{figure}
  \begin{center}
    \includegraphics*[width=0.45\textwidth]{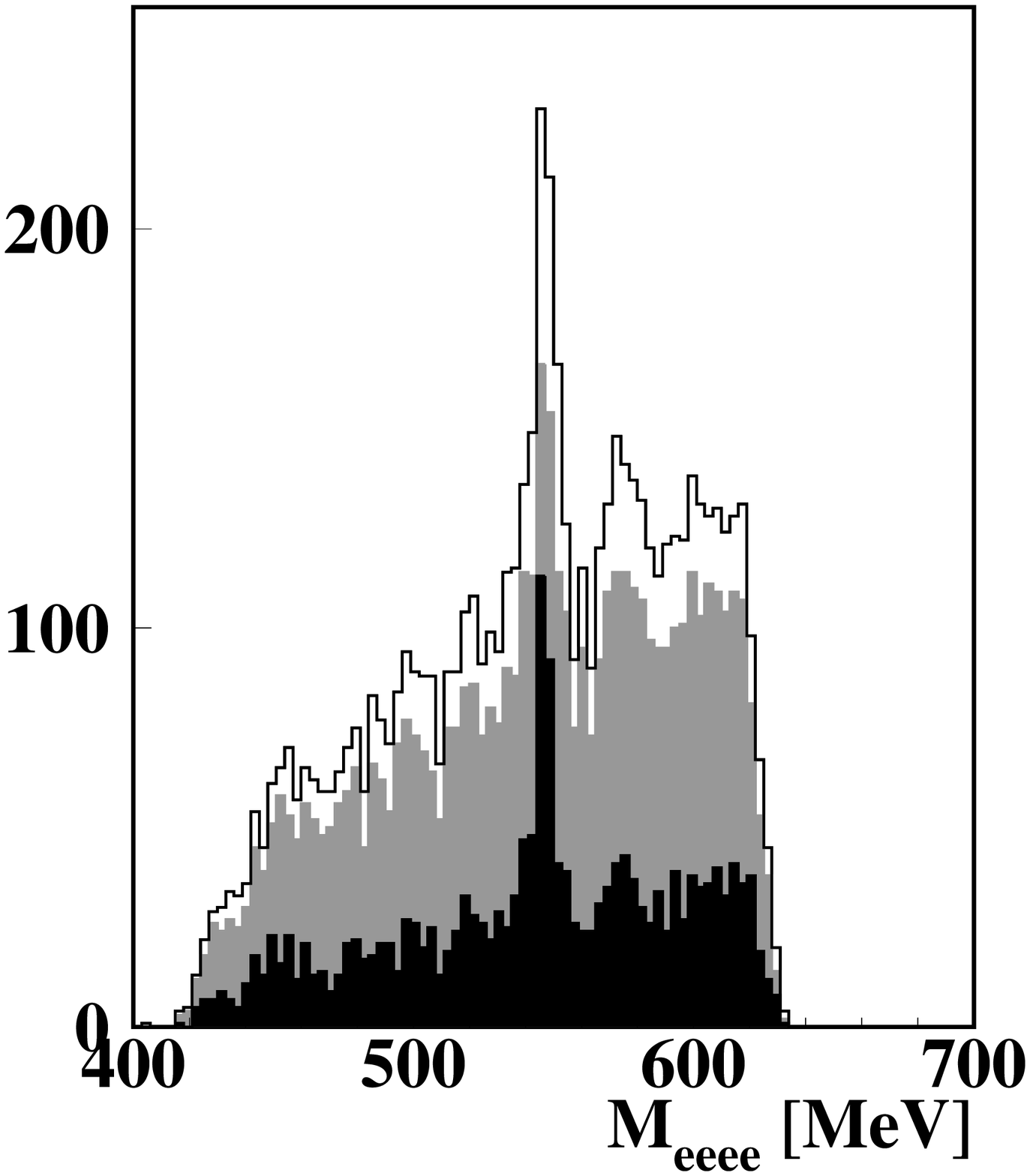}
    \includegraphics*[width=0.45\textwidth]{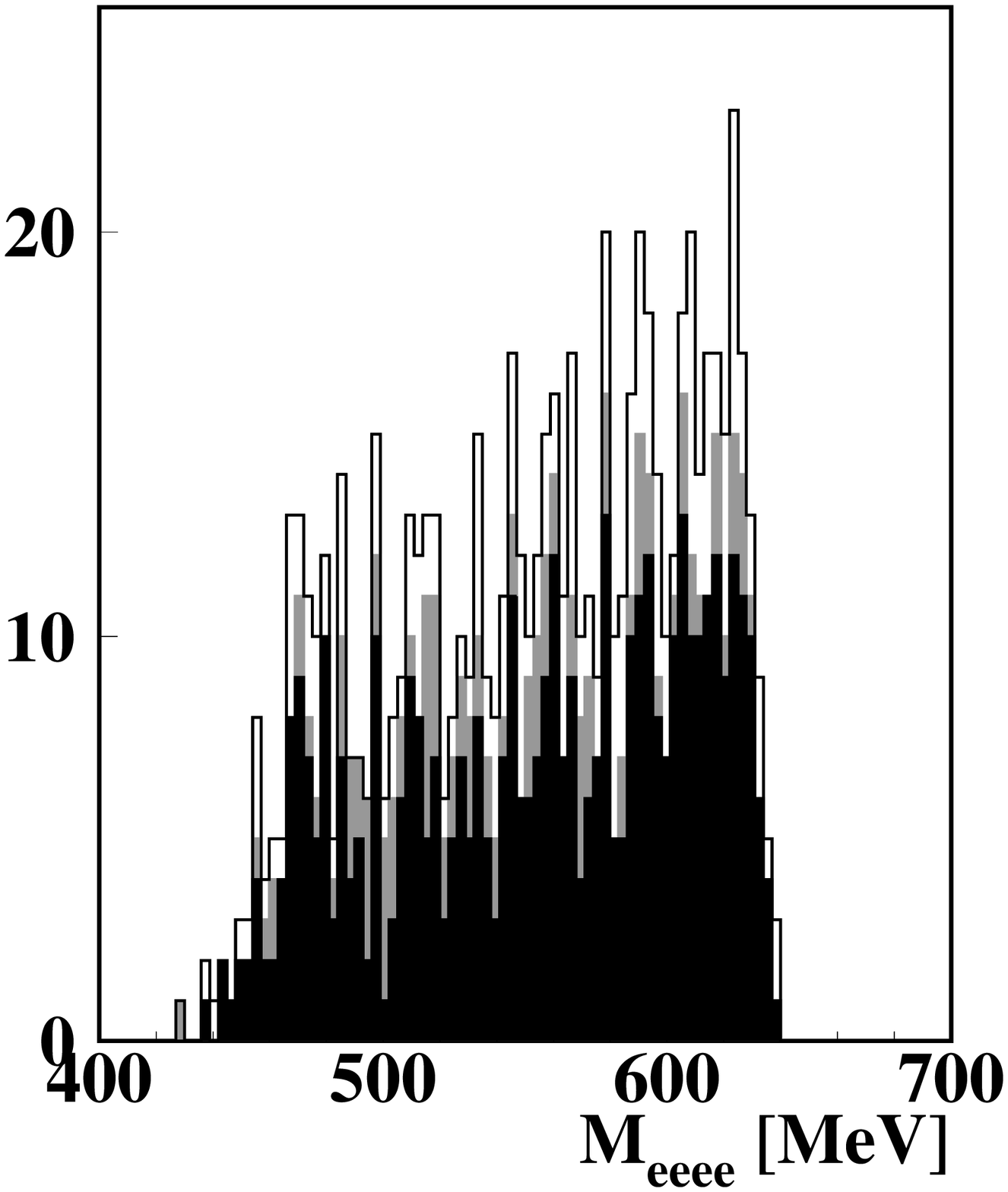}
    \includegraphics*[width=0.45\textwidth]{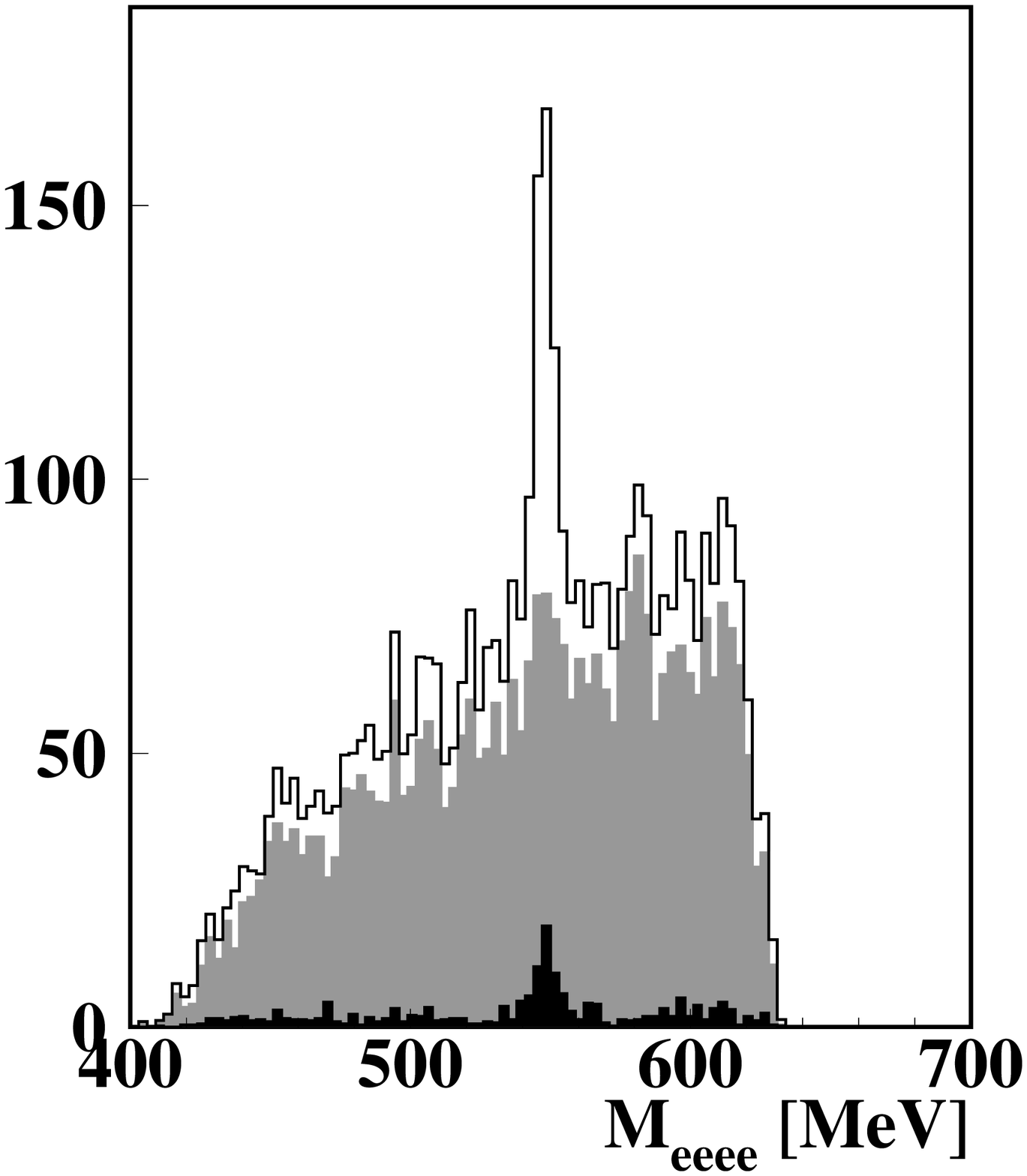}
    \includegraphics*[width=0.45\textwidth]{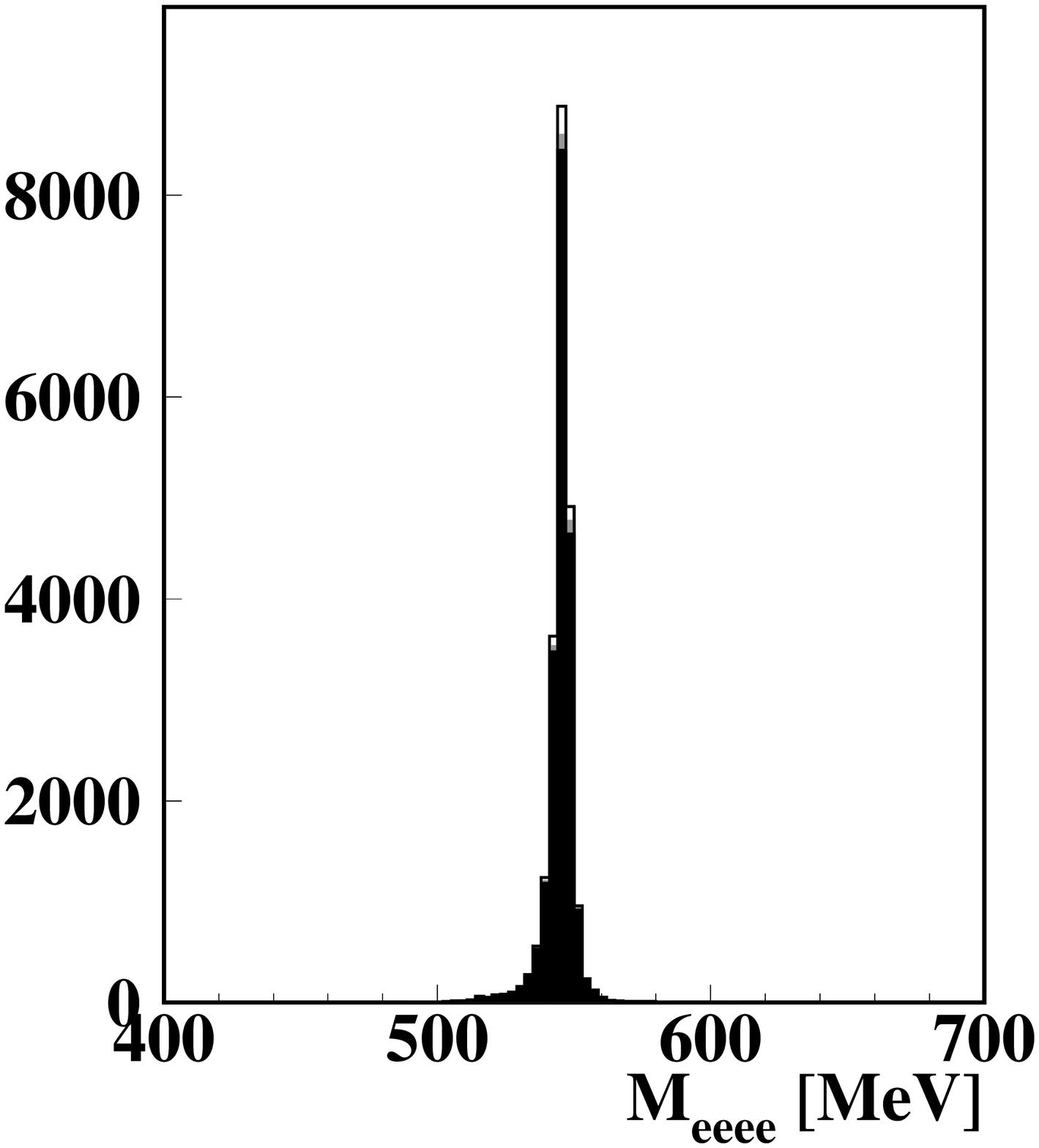}
  \end{center}
  \caption{$M_{\eeee}$ distribution after different analysis cuts:
           white: after the $\sfp$ and 
           the ${\langle \cos \theta \rangle}$ cuts;
           gray: after the cut on photon conversion;
           black: after the PID requirement.
           Top left: data; 
           top right: off-peak;
           bottom left: $\phi$ background Monte Carlo; 
           bottom right: signal Monte Carlo.
}
  \label{fig:cuts3}
\end{figure}

Two sources of background are present:
\begin{enumerate}
\item{$\phi$ background:
  \\
  this is mainly due to $\phippp$ events (with $\pi^0$ Dalitz decay)
  and to $\phietag$ events either with $\etappp$ (with $\pi^0$ Dalitz
  decay) or $\etappee$ or with $\etaeeg$ (with photon conversion on
  the Beam Pipe, BP, or the DC inner Wall, DCW).
  This last background has the same signature of the signal.
  Background from $\phikk$ is also present at the preselection level.
  \\
}
\item{$\ee$ continuum background:
  \\
  this is mainly due to $\bharad$ events with photon conversions,
  split tracks or interactions in the \dafne low beta quadrupoles.
  This background has been studied using off-peak data taken
  at $\sqrt{s}=1$ GeV, where $\phi$ decays are negligible. 
}
\end{enumerate}
A first background rejection is performed cutting on the sum of the
absolute value of the momenta of the four selected tracks requiring
$(600 < \sfp < 700)$ MeV.
\\
To remove $\ee$ continuum background from interactions in the low beta
quadrupoles, the quantities $\ltf$ and $\ltb$ have been defined as
the average polar angle of forward and backward selected particles.
Events having  $\ltf > 0.85$ and $\ltb < -0.85$ are rejected.
This cut has no effect on signal selection efficiency.
\\
To reject events due to photon conversion, each track is
extrapolated backward to the intersection with the BP and with the
DCW.
For each track pair, the invariant mass ($M_{ee}$) and 
the relative distance ($D_{ee}$) are computed.
A clear signal of photon conversion is visible in the
$D_{ee}$-$M_{ee}$ 2D plot for BP and DCW (figure \ref{fig:gee}).
Events having at least one combination satisfying 
$M_{ee}(BP) < 10$ MeV and $D_{ee}(BP) < 2$ cm or
$M_{ee}(DCW) < 30$ MeV and $D_{ee}(DCW) < 2$ cm are rejected.
\\
The last rejection is based on the Particle IDentification (PID) of
charged particles.
For each track associated to a calorimeter cluster, the quantity 
$\Delta t = t_{track} - t_{cluster}$ in both electron 
and pion hypothesis is evaluated; 
$t_{track}$ is defined as the length of the track divided by 
$\beta(m) c$.
Track with $\Delta t_e / \Delta t_{\pi} < 1 (>1)$ are identified as
electron (pion).
Events having more than two pions or no electrons are discarded.
\\
The effects of background rejection cuts on the various data
components are visible in figure \ref{fig:cuts3}, where the four
electrons invariant mass, $M_{\eeee}$, is shown at different steps
of the analysis.
In table~\ref{tab:Nev}, number of events in data, N(data), MC signal 
efficiencies, $\rm\varepsilon_{ana}(sig)$, and background rejection 
factor $R$, defined as the ratio of analysis efficiency between signal 
and background, are also reported. The $R$ value has been evaluated 
for three different cathegories: $\phietag$ with 
$\etaeeg$ ($R_{\etaeeg}$), $\phikk$ and $\phirpi$ ($R_{\phikkrpi}$)
and all other $\phi$ decays products ($R_{\rm others}$). 
After all cuts, background from kaons and $\phippp$ events is
negligible. The same holds for all other $\phi$ decays but $\etaeeg$
which, as will be shown in the next section, results in $\sim 15\%$
contamination level.
\begin{table}[t]
  \caption{Number of events in data, MC signal efficiency, background 
    rejection factor at different steps of the analysis.}
  \begin{center}
    \begin{tabular}{cccccr}
      \hline
      Cut & N(data) & $\rm\varepsilon_{ana}(sig)$& 
      $R_{\etaeeg}$ & $R_{\phikkrpi}$ &   $R_{\rm others}$ \\ 
      \hline
      Preselection                       & 451924 & $0.285(1)$ & 
      $1.86(2)\times 10^{2}$   & $5.01(2)\times 10^{3}$ & 
      $1.435(8)\times 10^{3}$ \\
      $\chi^2$                           &  36282 & $0.217(1)$ &
      $2.01(3)\times 10^{2}$   & $1.13(1)\times 10^{5}$ &
      $3.44(5)\times 10^{4}$ \\
      $\Sigma_1^4|\vec{p}_i|$            &  16811 & $0.216(1)$ &
      $2.68(5)\times 10^{2}$   & $2.21(3)\times 10^{5}$ &
      $\ 6.9(1)\times 10^{4}$ \\
      $\cos\theta_b$, $\cos\theta_f$  &  15003 & $0.216(1)$ &
      $2.68(5)\times 10^{2}$   & $2.21(3)\times 10^{5}$ &
      $\ 6.9(1)\times 10^{4}$ \\
      $\gamma$ conversion               &  12198 & $0.209(1)$ &
      $1.11(4)\times 10^{3}$ & $2.53(3)\times 10^{5}$ &
      $1.13(3)\times 10^{5}$ \\
      PID                                &   4239 & $0.205(1)$ &
      $1.12(4)\times 10^{3}$ & $1.02(8)\times 10^{7}$ &
      $\ 5.1(3)\times 10^{5}$ \\ 
      \hline
    \end{tabular}
  \end{center}
  \label{tab:Nev}
\end{table}
Systematics on the Monte Carlo description of photon conversion have 
been studied using events with similar characteristics. 
A clean control sample is provided by the $\phietaee$, 
$\etappp$ decay chain, where simple analysis cuts provide 
a good data-MC agreement, with negligible background contamination. As 
for the $\etaeeee$ channel, before dedicated analysis 
cuts the control sample is significantly contaminated by background from 
photon conversion ($\phietag$ with photon converting on beam 
pipe or drift chamber walls). This background is completely removed 
rejecting events with $D_{ee}(DCW) < 10$ cm and $M_{ee}(DCW) < 80$ MeV.
For the $\etaeeee$ channel this cut has not been applied 
because, having two electrons and two positrons in the final state, 
the search for a conversion has to be performed over all the four 
$\ee$ combinations, thus spreading the signal contribution in the 
$M_{ee}(DCW)$--$D_{ee}(DCW)$ plane and lowering significantly the 
analysis efficiency.
Removing the cuts on $M_{ee}$--$D_{ee}$ planes in the control sample,
a clear background contamination from photon conversion is visible. 
Data-MC comparison shows that, increasing in the simulation the 
probability of conversion by 10\%, an excellent agreement is found. 
A 10\% systematic error is then assigned to photon conversion 
and added to the uncertainties coming from MC statistics and 
BR($\etaeeg$) measurement \cite{pdg2010}:
$N(\etaeeg) = 80 \pm 3_{\rm MC} \pm 8_{\rm BR} \pm 8_{\rm syst}$.


\section{Evaluation of the 
  BR(\mathversion{bold}$\etaeeeeg$\mathversion{normal})}
\label{sec:fit-invmass}

\begin{figure}
  \begin{center}
    \includegraphics*[width=0.49\textwidth]{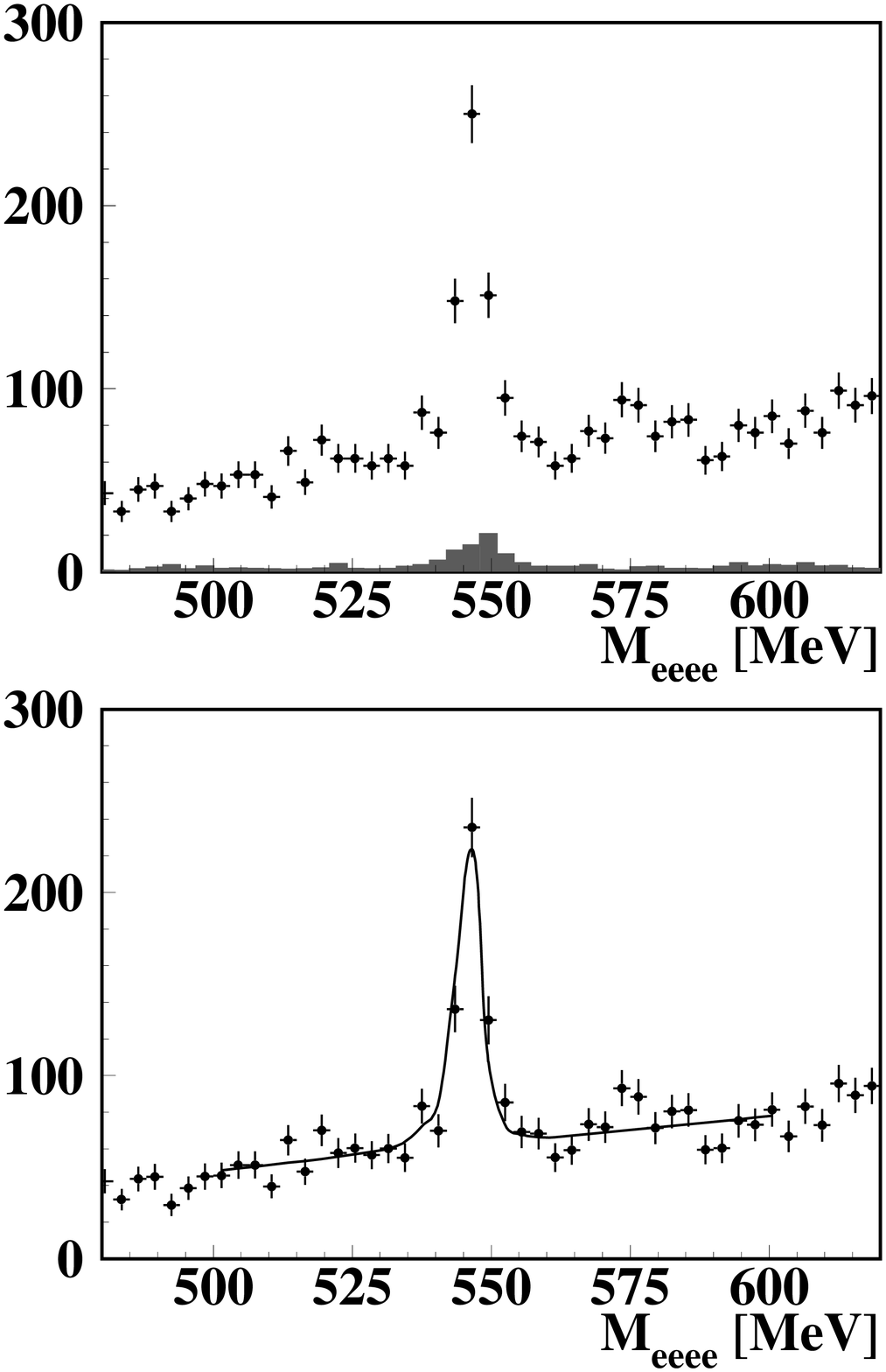}
  \end{center}
  \caption{Top panel: $M_{\eeee}$ data distribution at the end of the
           analysis chain; the expected $\phi$ background MC shape is
           shown in dark gray.
           Bottom panel: $M_{\eeee}$ fit to data after $\phi$ background
           subtraction.
}
  \label{fig:fit}
\end{figure}

As discussed in the previous section, the only significant background 
contamination surviving all the analysis cuts is due to $\etaeeg$ events
with photon conversion, that have a signature similar to the signal.
The overall estimated background from $\phi$ decays has been subtracted 
bin-by-bin to the $M_{\eeee}$ spectrum obtained in data (figure 
\ref{fig:fit} top), taking into account also systematic errors. 
The event counting is done fitting the resulting spectrum with the two 
residual contributions: signal and $\ee$ continuum background events. 
The $M_{\eeee}$ shape for the signal is obtained by fitting MC events with
two Gaussian functions plus a third order polynomial function.
The fit range is $500<M_{\eeee}<600$ MeV.
The $M_{\eeee}$ distribution for $\ee$ continuum events has been studied
on the data taken at $\sqrt{s} = 1$ GeV, where contributions from
$\phi$ decays are suppressed. 
Even though the small statistics of the sample does not allow to
precisely extract the shape, a first order polynomial well reproduces
the data in the signal region.
The free parameters are an overall scale factor for signal and the two
parameters describing the $\ee$ continuum background. 
Fit results are shown in figure \ref{fig:fit} bottom.
The resulting $\chi^2/{\rm ndf}$ is 43.9/34, corresponding to 
P$(\chi^2)=0.12$. 
The number of signal events is $N(\etaeeee) = 362 \pm 29$.
The branching ratio has been evaluated according to the formula:
\begin{equation}
  BR(\etaeeeeg) = \frac{N_{\etaeeeeg}}{N_{\eta\gamma}} \cdot
                 \frac{1}{\epsilon_{\etaeeeeg}}
   \label{eq:masterformula}
\end{equation}
where $N_{\etaeeeeg}$ is the number of signal events and 
$\epsilon_{\etaeeeeg}$ is the efficiency taken from MC. 
The number of $\phietag$ events, $N_{\eta\gamma}$, has been obtained
using the formula $N_{\eta\gamma} = \mathcal{L} \cdot \sigma_{\phietag}$, 
where $\mathcal{L}$ is the integrated luminosity and the cross section
$\sigma_{\phietag}$ has been evaluated taking into account the $\phi$
meson line shape on a run by run basis \cite{csnote}.
Inserting all the numbers quoted in table \ref{tab:summary},
the value:
\begin{equation}
  BR(\etaeeeeg) = (2.44 \pm 0.19_{stat+bckg}) \times 10^{-5}
\end{equation}
is obtained, where the error accounts for the uncertainty of the fit
result.
\begin{table}
  \begin{center}
    \caption{Summary of the numbers used in the master formula
             (\ref{eq:masterformula}) for the branching ratio evaluation.}
    \label{tab:summary}
    \begin{tabular}{cc}
      \hline
      BR inputs                       & Values\\
      \hline                     
      Number of events                & $362 \pm 29$ \\
      Efficiency $\epsilon_{\etaeeeeg}$ & $0.205 \pm 0.001$ \\
      Luminosity                      & $(1733 \pm 10)\ {\rm pb}^{-1}$ \\
      $\eephietag$ cross section      & $(41.7 \pm 0.6)\ {\rm nb}$ \\
      \hline
    \end{tabular}
  \end{center}
\end{table}
\\
The systematic uncertainties due to analysis cuts have been evaluated
by applying separately a variation of $\pm1\,\sigma$ on all variables 
and re-evaluating the branching ratio. 
The $\sigma$ values have been obtained using MC signal events.
For the $\chi^2$ variable the cut has been moved by $\pm 500$, while 
the particle identification cut has been changed by $\pm 10\%$. 
The systematic error on the fit to the $M_{\eeee}$ distribution has been
evaluated considering:
\begin{itemize}
\item the binning of the $M_{\eeee}$ histogram, changed from
  3 MeV, used as default, to 2 and 4 MeV;
\item the $M_{\eeee}$ range, enlarged and reduced by 10 MeV on both sides;
\item the slope of the $\ee$ continuum background has been fixed to the
  value obtained from off-peak data fit.
\end{itemize}
The relative variation of the BR for each source of systematic
uncertainty is reported in table \ref{tab:systematics}.
The uncertainty on $N_{\eta\gamma}$ has been added to the systematics
in the normalization term.
The total error is taken as the quadratic sum of all contributions.
\begin{table}[!t]
  \begin{center}
    \caption{Summary table of systematic uncertainties.}
    \begin{tabular}{cc}
      \hline
      Source of uncertainty    & Relative error \\
      \hline
      $\chi^2$                 & $-0.51\%\ / +2.62\%$ \\
      $\ltb$ and $\ltf$        & $-0.04\%\ / +0.47\%$ \\
      $\sfp$                    & $+0.11\%$ \\
      $\gamma$ conversion      & $-0.74\%\ / +2.43\%$ \\
      PID                      & $+1.84\%$\\
      Fit range                & $-0.38\%\ / +1.13\%$ \\
      Binning on $M_{\eeee}$    & $-3.21\%\ / +0.19\%$ \\
      Background slope         & $+0.38\%$ \\
      Normalization            & $\pm 1.64\%$ \\
      \hline	     		  
      Total                    & $-3.73\%\ / +4.53\%$\\
      \hline    
    \end{tabular}
    \label{tab:systematics}
  \end{center}
\end{table}

\section{Conclusions}
\label{sec:conclusions}
Using a sample of 1.7 fb$^{-1}$ collected in the $\phi$ meson mass
region, the first observation of the $\etaeeeeg$ decay has been
obtained on the basis of $362 \pm 29$ events.
The corresponding branching ratio is:
\begin{equation}
  BR(\etaeeeeg) = 
  ( 2.4 \pm 0.2_{\rm stat+bckg} \pm 0.1_{\rm syst}) \times 10^{-5} \, .
\end{equation}
Radiative events slightly modify momentum distribution of the charged
particles and have been carefully considered in the efficiency evaluation.
As a result, the measured branching ratio is fully radiation inclusive.
\\
Our measurement is in agreement with theoretical predictions, which
are in the range $( 2.41 - 2.67) \times 10^{-5}$ 
\cite{Jarlskog,Miyazaki:1974qi,Bijnens:1999jp,Lih:2009np,Petri:2010ea}.

\section*{Acknowledgments}
\label{acknowledgments}
We would like to thank J.~Bijnens for the useful discussions and for
having provided the signal Monte Carlo generator.
We thank the DAFNE team for their efforts in maintaining low background
running conditions and their collaboration during all data-taking. 
We want to thank our technical staff: 
G.~F.~Fortugno and F.~Sborzacchi for their dedication in ensuring
efficient operation of the KLOE computing facilities;
M.~Anelli for his continuous attention to the gas system and detector
safety;
A.~Balla, M.~Gatta, G.~Corradi and G.~Papalino for electronics
maintenance;
M.~Santoni, G.~Paoluzzi and R.~Rosellini for general detector support;
C.~Piscitelli for his help during major maintenance periods.
This work was supported in part by EURODAPHNE, contract FMRX-CT98-0169; 
by the German Federal Ministry of Education and Research (BMBF) contract
06-KA-957; 
by the German Research Foundation (DFG), 'Emmy Noether Programme', 
contracts DE839/1-4;
by the EU Integrated Infrastructure Initiative HadronPhysics Project
under contract number RII3-CT-2004-506078;
by the European Commission under the 7th Framework Programme through
the 'Research Infrastructures' action of the 'Capacities' Programme,
Call: FP7-INFRASTRUCTURES-2008-1, Grant Agreement N. 227431;
by the Polish Ministery of Science and Higher Education through the
Grant No. 0469/B/H03/2009/37.




\end{document}